\documentclass{an}
\usepackage{times}
\usepackage{graphicx}
\begin{document}%


\newcommand{\xmmn}{{\it XMM-Newton\/}}
\newcommand{\chan}{{\it Chandra\/}}
\newcommand{\hst}{{\it HST\/}}

\def\kevper{{\rm ~keV~cm^{-2}~s^{-1}~sr^{-1}~keV^{-1}}}
\def\Lsun{\hbox{$\rm ~L_{\odot}$}}
\def\Msun{\hbox{$ M_{\odot}$}}
\def\ergcms{{\rm ~erg~cm^{-2}~s^{-1}}}
\def\ergams{{\rm ~erg~arcmin^{-2}~s^{-1}}}
\def\ergsec{{\rm ~erg~s^{-1}}}
\def\chisq{{$\chi^{2}$}}
\def\rchi{{$\chi^{2}_{\nu}$}}
\def\atpcm{{\rm ~atoms~cm^{-2}}}
\def\ctsec{{\rm ~count~s^{-1}}}
\def\dg{^{\circ}}
\def\H0{{\rm ~km~s^{-1}~Mpc^{-1}}}
\def\ctsecam{{\rm ~count~s^{-1}~arcmin^{-2}}}
\def\snow{{\times 10^{-6} \rm ~count~s^{-1}~arcmin^{-2}}}
\def\ctpix{{\rm ~count~pixel^{-1}}}
\def\cext{{C_{\rm ext}}}
\def\xspnorm{{\rm ~photon~cm^{-2}~s^{-1}~keV^{-1}}}

\def\etal{et al.~\/}
\def\eg{{\it e.g.,~\/}}
\def\etc{{\it etc.~\/}}
\def\ie{{\it i.e.,~\/}}
\def\cf{{\it cf.~\/}}
\def\la{\mathrel{\hbox{\rlap{\hbox{\lower4pt\hbox{$\sim$}}}{\raise2pt\hbox{$<$}}
}}}
\def\ga{\mathrel{\hbox{\rlap{\hbox{\lower4pt\hbox{$\sim$}}}{\raise2pt\hbox{$>$}}
}}}
\def\d25{$D_{25}$}
\def\nh{{$N_{\rm H}$}}
\def\Ha{{H$\alpha$}}
\def\Hb{{H$\beta$}}
\def\hi{H{\small I}$~$}
\def\hii{H{\small II}$~$}
\def\lx{$L_{\rm X}$}
\def\fx{$f_{\rm X}$}
\def\deg{\hbox{$^\circ$~\/}}
\def\arcm{\hbox{$^\prime$~\/}}
\def\arcs{\hbox{$^{\prime\prime}$~\/}}

\title{Key results from an {\em XMM-Newton}~and {\em Chandra}~study of a new sample of extreme ULXs from the 2XMM catalogue}

\author{A. D. Sutton\inst{1} \fnmsep\thanks{\email{andrew.sutton@durham.ac.uk}}
\and T. P. Roberts\inst{1}
\and D. J. Walton\inst{2}}

\titlerunning{A new sample of extreme ULXs}
\authorrunning{A. D. Sutton, T. P. Roberts \& D. J. Walton}
\institute{Department of Physics, University of Durham, South Road, 
Durham, DH1 3LE, UK
\and
Institute of Astronomy, University of Cambridge, Madingley Road, 
Cambridge, CB3 0HA, UK}



\abstract{We present highlights from a study of a sample of 10 
extreme-luminosity candidate ultraluminous 
X-ray sources ($L_{\rm X}>5\times10^{40}~\ergsec$), all at distances $<$ 100 
Mpc, identified from a cross-correlation of the RC3 catalogue of galaxies with 
the 2XMM catalogue.  Five of the sample have also been observed by \chan.  Of 
the 10 sources, seven reside in the disc or arms of spiral galaxies, and the 
remaining three are close to large elliptical galaxies.  
Unlike many less luminous ultraluminous X-ray sources, 
temporal variability is observed on short (ks) and long (year) timescales for most sources
in our sample. 
Long term spectral variability is also evident 
in some sources.  In one case, we use archival \chan~data to demonstrate that 
a hyperluminous X-ray source candidate identified by \xmmn~is actually resolved
into multiple point sources at high spatial resolution, but note that the 
other candidates remain unresolved under \chan's~intense scrutiny.}

\maketitle

\section{Introduction}

Ultraluminous X-ray  sources (ULXs) are extra-nuclear X-ray sources with 
luminosities in excess of the Eddington limit of a stellar mass black hole 
($L_{\rm X} > 10^{39}~\ergsec$). The physical nature of these intriguing objects
remains open to debate (\eg Miller \& Colbert 2004, Roberts 2007).
It was suggested by Colbert \& Mushotzky (1999) that ULXs were powered by
accretion on to intermediate-mass black holes (IMBHs) of mass $10^2$--$10^4
~\Msun$. This suggestion was supported by the detection of seemingly
cool accretion disc components in the spectra of a number of ULXs, with 
apparent temperatures of $\sim$ 0.1--0.3 keV indicating $\sim~1000~\Msun$ 
black holes.

However the association of large numbers of ULXs with regions of rapid star
formation (\eg Fabbiano et al. 2001; Gao et al. 2003) implies that they are short-lived,
hence require enough progenitors for multiple generations to be observed. The required
number density would imply that an unfeasible proportion of galaxy mass would end up in IMBHs 
so it is much more likely that most ULXs harbour
stellar remnant black holes (King 2004) and are either subject to beaming (geometric, King 
et al. 2001; relativistic, K{\"o}rding et al. 2002) or accreting at super-Eddington 
rates.

Re-analyses by Gladstone et al. (2009), of a  sample 
of ULX X-ray spectra chosen only
on the basis of very high data quality ($\ge$ 10000 EPIC counts) have also brought into 
question the IMBH interpretation, instead identifying them with a previously unknown, 
presumably super-Eddington ``ultraluminous'' accretion state. 
This is consistent with the consensus view that the majority of ULXs are increasingly
likely to be powered by accretion onto small (up to 100 $\Msun$) black
holes (\eg Roberts 2007).

The above arguments do not preclude the possibility that a minority of ULXs are
indeed powered by accretion onto IMBHs. Possibly the best IMBH candidates
are the most luminous ULXs, with extraordinary X-ray luminosities in excess of
$5 \times 10^{40}~\ergsec$, including the hyperluminous X-ray sources (HLXs) 
with $L_{\rm X} > 10^{41}~\ergsec$. These extreme sources sit above the steep turn-off 
in the X-ray luminosity function of extra-nuclear sources (Grimm et al. 
2003), below which sources can be explained by super-Edd\-ington accretion rates onto
stellar-mass black holes (King 2008) or larger (up to 100 $\Msun$) stellar
remnants (Zampieri \& Roberts 2009). 
Sources above the break require a combination of both larger black holes 
and super-Eddington accretion rates; or perhaps they harbour the elusive
IMBHs. In-order to investigate such a possibility, we present key results from a study of a 
new, small sample of some of the most extreme luminosity ULXs observed by \xmmn~and \chan.

\section{The sample of extreme ULXs}

\begin{table*}
\caption{ULX sample}
\begin{tabular}{ccccccc}
\hline
ID & 2XMM Source & Host Galaxy$^a$ & Galaxy Type$^b$ & Separation $^c$ & Distance$^d$ & Peak Luminosity$^e$
\\
&&&& $^{\prime \prime}$ & Mpc & $10^{40}$ erg s$^{-1}$
\\
\hline
Src. 1 & 2XMM J011942.7+032421 & NGC 470 & SA(rs)b & 33 & 32.7 & $10.3^{+0.7}_{-1}$
\\
Src. 2 & 2XMM J024025.6-082428 & NGC 1042 & SAB(rs)cd & 96 & 18.9 & $3.6^{+0.1}_{-0.2}$
\\
Src. 3 & 2XMM J072647.9+854550 & NGC 2276 & SAB(rs)c & 45 & 33.3 & $6.1^{+0.4}_{-0.6}$
\\
Src. 4 & 2XMM J120405.8+201345 & NGC 4065 & E & 21 & 88 & $12 \pm 3$
\\
Src. 5 & 2XMM J121856.1+142419 & NGC 4254 & SA(s)c & 103 & 33.2 & $8.9^{+0.7}_{-2}$
\\
Src. 6 & 2XMM J125939.8+275718 & NGC 4874 & cD0 & 57 & 99.8 & $20 \pm 4$
\\
Src. 7 & 2XMM J134404.1-271410 & IC 4320 & S0? & 18 & 95.1 & $27 \pm 3$
\\
Src. 8 & 2XMM J151558.6+561810 & NGC 5907 & SA(s)c: edge-on & 102 & 14.9 & $4.2 \pm 0.1$
\\
Src. 9 & 2XMM J163614.0+661410 & MCG 11-20-19 & Sa & 16 & 96.2 & $7^{+2}_{-1}$
\\
Src. 10 & 2XMM J230457.6+122028 & NGC 7479 & SB(s)c & 68 & 32.8 & $6.1^{+0.3}_{-0.4}$
\\
\hline
\end{tabular}
\\Notes:
$^a$ Galaxy with which the candidate ULX was initially identified by the cross
correlation.
$^b$ Galaxy morphology from de Vaucouleurs et al. (1991).
$^c$ Galaxy centre - ULX candidate angular separation (Walton et al., in prep.).
$^d$ Galaxy distance used in the analysis. With the exception of src. 8,
cosmology corrected distances based on redshifts from de Vaucouleurs et al. (1991)
were used. For src 8., due to the requirement for local corrections the distance 
from Tully (1988) was used.
$^e$ Peak observed 0.3 -- 10 keV luminosity of the ULX candidates if at the identified
host galaxy distance, based on spectral fits (see sec. 3.2).
\label{obs}
\end{table*}

The extreme ULX sample was retrieved from a catalogue of 475 ULX candidates in 
240 galaxies (Walton et al. in prep.), produced by cross-correlating 
the RC3 catalogue of galaxies
(de Vaucouleurs et al. 1991) with the 2XMM DR1 catalogue (Watson et al. 2009).
From this catalogue twelve 
candidate 
ULXs observed by \xmmn, 
with $L_X > 5 \times 10^{40}~{\rm erg~s^{-1}}$ (based on the catalogued 2XMM flux) 
and within a distance of 100 Mpc 
were identified. The sample was reduced to 
ten sources by the exclusion of M82 X-1 as it has been previouly well studied 
and a probable spurious detection in NGC 4889 where it was unclear whether significant
source counts were detected in excess of the clumpy galactic ISM.

Of the ten bright ULX candidates observed by \xmmn, and listed in table \ref{obs}, 
three sources were 
previously identified as ULXs (src. 2, 3 and 5), 
and eight were present in multiple observations (the exceptions being src. 7 and 10 
which were detected in only one \xmmn~observation), including five 
sources detected by \chan~(src. 1, 3, 5, 6 and 9). Six  of the sources 
were in the arms of spiral galaxies, one
additional source was in an edge-on disc, and the remaining three sources were 
near to large elliptical galaxies (see table \ref{obs}). The catalogued luminosities of 
four sources (src. 1, 4, 6, 7),
including all three of the ULX candidates associated with elliptical galaxies,
were consistent with HLXs \footnote{Src. 3 was previously reported as a HLX (Davis \& Mushotzky
2009) but the distance used here implies a lower luminosity.}.

\section{Analysis and results}

We detail the data reduction, and provide more detail on the analysis, in
Sutton \etal (in prep.). Here we highlight some interesting results.

\begin{figure}
\includegraphics[width=8.cm]{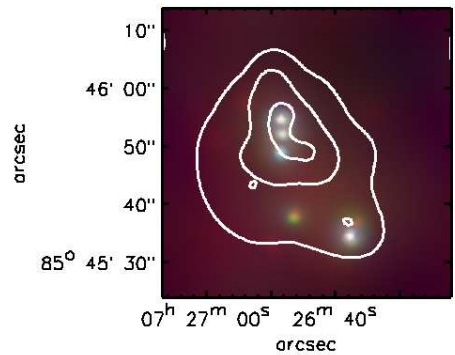}
\caption{Smoothed three colour \chan~image of the \xmmn~region of src. 3, the candidate
extreme ULX in NGC 2276, where the rgb colours correspond to 0.2--1.5, 1.5--2.5 and 
2.5--8 keV. The image is overlaid with \xmmn~contours, 
calculated 
from the smoothed \xmmn~image of the source. Contours represent photon count 
levels of $10^0$, $10^{-0.5}$ and $10^{-1}$${\rm ~ct~pix}^{-1}$.}
\label{ngc_2276}
\end{figure}

Four out of the five 2XMM ULX candidates imaged by \chan~remain point-like in 
appearance at higher spatial resolution (src. 1, 5, 6 and 9); however the object reported
by Davis \& Mushotzky (2004), here src. 3, is resolved into multiple point 
sources (Fig. \ref{ngc_2276}). If at the distance of NGC 2276, all three sources are 
luminous enough to be candidate ULXs, although their total flux is less than 
that previously observed for the unresolved source by \xmmn. Clearly at least 
one of these sources was substantially more luminous during the earlier 
observation. The fact that short-term variability was detected in the 
unresolved source (\cf Fig. \ref{Fvar}) suggests that its flux was dominated 
by a single source.

Initial estimates from Walton et al. (in prep.) 
were for at most $\sim$ 1 in 4 candidates with $L_x > 5 
\times 10^{40}~\rm{erg~s^{-1}}$ to be 
background contaminants. For two of the sources 
candidate counterparts were identified
implying that they are possible foreground or background contaminants (see below). 
Both candidate contaminant sources were 
initially 
associated with elliptical galaxies and were amongst the highest implied 
luminosity sources in our sample. Src. 7, in 
the elliptical galaxy IC 4320, remains as the most luminous detection and only 
object in an elliptical in the sample.
 
The \xmmn~error region of src. 4, the HLX candidate initially 
identified as being associated with NGC 4065, was shown to be coincident with 
SDSS J120405.84+2013\-45.1 (star, Adelman-McCarthy et al. 2008) 
and SDSS J12\-04\-05.83+201345.0 (QSO, Schneider et al. 2010). 
The 2--7.5 keV flux of src. 4 ($\sim 3.74 \times 10^{-14}~{\rm erg~cm^{-2}~s^{-1}}$)
and R band magnitude of SDSS J120405.83+201345.0 ($\sim 19.8$, Schneider et al. 2010) 
combine to give an X-ray to optical flux ratio of $\sim$ 1 consistent with an AGN 
(\cf Fig. 11 of 
Caccianiga et al. 2008) with the caveat that the spectrum and flux estimate of src. 4 
may be contaminated by the nearby faint X-ray source 2XMM J120406.1+201406. 
We tentatively identify this HLX candidate as a contaminant, although {\chan} data is 
required to confirm this. 

The HLX candidate src. 6 was initially identified with NGC 4874. 
The \xmmn~error region showed it may instead be associated with the smaller 
satellite galaxy SDSS J1259\-39.65+\-275714.0. This was confirmed by a later 
\chan~observation (obs ID 10612), and its location $\sim$ 3 arcsecs from the centre of 
the satellite galaxy maintained it as a good HLX candidate.  However, an optical 
point source was identified, using 
{\it Hubble Space Telescope}~(\hst) Advanced Camera for Surveys/Wide Field Channel 
archived data, at the position 
of the HLX candidate. The source was therefore conservatively 
excluded as a possible contaminant, however we note that this may be revised after 
further examination of the \hst~data.

\subsection{Variability}

\begin{figure}
\includegraphics[width=8.cm]{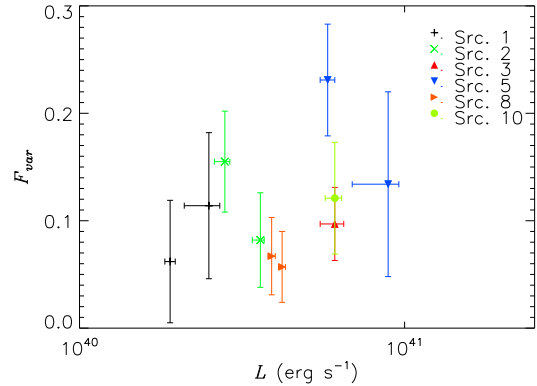}
\caption{Excess fractional variability of ULX candidates in which variability
was detected at greater than 1$\sigma$ significance.}
\label{Fvar}
\end{figure}

A characteristic of ULXs as a class is that they show little short-term 
variability -- for example Swartz et al. (2004) found that only 5-15\% of ULX 
candidates displayed detectable variability on time scales of 1 ks, 
and  Heil et al. (2009) showed 
that variability is suppressed in a number of high quality ULX datasets in the $10^{-4}$ --
1 Hz frequency range, many of which are limited by lack of statistics or lack of 
monitoring observations. 
To test for variability, light curves of the sample ULXs were binned on time scales 
such there were $\sim$ 25 counts per element of temporal resolution. 
Fractional variability (calculated using the method of Vaughan \etal 2003) was 
detected in a total of 10 detections of 6 sources out of the 8 
remaining sources (Fig. \ref{Fvar}) albeit mainly at low significance. 
Similar levels of variability could not be ruled out in 
the remaining two sources. If variability is common amongst 
extreme 
ULXs, we could be observing variability in excess of that expected
from a linear rms variability -- flux relation (\cf Heil \& Vaughan 2010).

\begin{figure}
\includegraphics[width=8.cm]{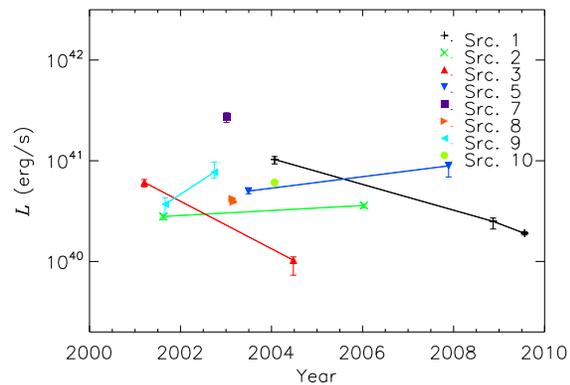}
\caption{0.3--10 keV \xmmn~\&~\chan~absorbed luminosities of ULX candidates
(excepting those that have been excluded as possible contaminant objects)
for observations with greater than 100 counts. For the later 
observation of src. 3 the brightest of the resolved point sources is plotted.}
\label{date_L}
\end{figure}

Src. 1, 3 \& 5 were present in multiple observations and were observed to vary 
in luminosity, by factors $\sim$ 1.5--7 between detections (Fig. \ref{date_L}). 
Interestingly, in a number of sources with 
multiple observations, the peak luminosity is not sustained, \eg src. 1 
decreases from the hyperluminous regime to a more typical high ULX luminosity.

\subsection{Spectral analysis}

Absorbed power-law and multi-colour-disc blackbody spectral 
models (powerlaw and disk\-BB in {\sc xspec}) 
were fitted to data with $> 250$ counts using $\chi^2$ statistics and $> 100$ counts using 
Cash-statistics (Cash 1979). Two absorption components 
were included, one fixed to the Galactic foreground column density
(Dickey \& Lockman 1990), the other 
free. A power-law model is not rejected at the 3$\sigma$ level in any observation. 
Power-law spectral fits had typical intrinsic absorption columns of 
$\sim$ 0.1-1$\times10^{22}~{\rm cm^{-2}}$ and generally fairly hard photon 
indexes of 1.5-2.2.  For perspective, typical values of spectral parameters derived from 
ULX samples over a more complete luminosity range are: 
$N_H \sim$ 0.09--0.57, 0.02--3 $\times 10^{22}~{\rm cm^{-2}}$ and 
$\Gamma \sim$ 1.6--3.3, 0.8--4 (Gladstone et al. 2009; Berghea et al. 2008).
Swartz et al. (2004) found $\langle \Gamma \rangle = 1.74 \pm 0.03$ for their sample
of \chan~ULX observations,
with a minority of sources (20 out of 130) having $\Gamma > 3$.
The sample presented here is interesting in that it appears to be consistent with the 
trend observed in the sample of Berghea et al. (2008), that the brightest ULXs
are spectrally harder than less luminous ULXs. 
 
Thermal (disc-dominated) spectra were statistically preferred (although power-law models
could not be rejected in any detection) in the most 
luminous observation of src. 1, in 2 of the 3 resolved sources in the 
later observation of src. 3, and the later observation of src. 8.  
Disc temperatures varied between $\sim$ 
1.0--1.6 keV, similar to temperatures seen 
in other luminous ULXs (when modelled as a multi-colour-disc, \eg 1.1 -- 1.8 keV,
Makishima et al. 2000)
and in galactic black hole binaries in the thermal dominated state. 
However, with the moderate to low quality X-ray data available it might not be 
possible to discriminate between a thermal and ultraluminous state model
(Gladstone \& Roberts 2009).
 
\begin{figure}
\includegraphics[height=8.cm, angle=-90]{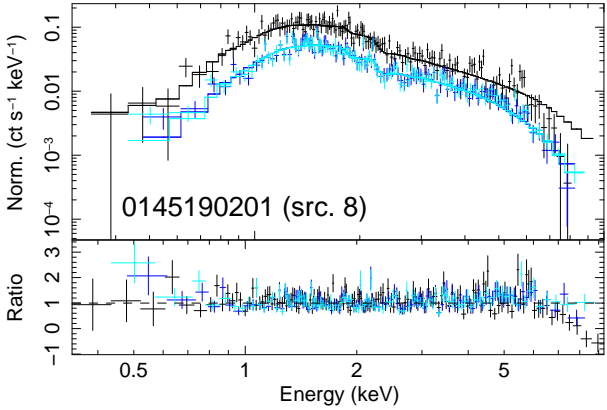}
\caption{Power-law fit to \xmmn~EPIC detection of src. 8, 
the extreme ULX candidate in spiral galaxy
NGC 5907. A spectral turn-over at $\sim$ 6 keV is clearly 
visible.}
\label{spectra}
\end{figure}

Observations of src. 8 -- with clearly the best X-ray 
data in the sample -- contain significant evidence of a high energy spectral 
break (Fig. \ref{spectra}). Following the analysis of Stobbart \etal (2006) 
the high energy 2-10 keV spectrum (over which absorption is negligible) was fitted 
with both a standard and 
a broken power-law. The broken power-law gives a significant improvement in 
both observations (F-test probabilities of $1 \times 10^{-12}$ and 
$4 \times 10^{-8}$) with break energies of $5.9^{+0.4}_{-0.2}$ and $6.5^{+0.3}_{-0.4}$
keV for the first and second observations respectively, an observational signature of 
the ultraluminous state 
(Gladstone et al. 2009).  Thus it appears the ultraluminous state 
(\ie super-Eddington accretion) is still present in at least one ULX in the sample.
 
\begin{figure}
\includegraphics[width=8.cm]{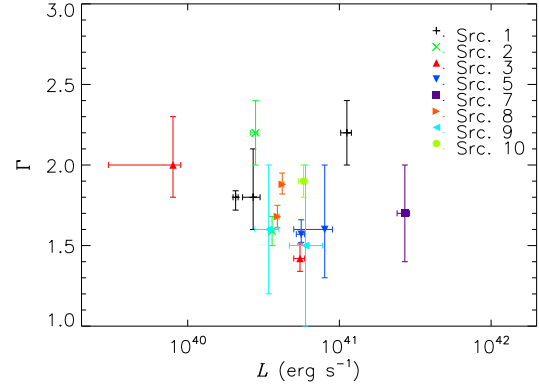}
\caption{Power-law photon index plotted against luminosity for
ULX candidate observations with greater than 100 counts. 90\% 
errors are shown for spectral index.}
\label{L_gamma}
\end{figure}

Variability in  spectral index was also
studied (Fig. \ref{L_gamma}). The sample appears to possess heterogeneous 
behaviours. The spectra of src. 1 and 8 are softer at 
increased luminosities, whereas that of src. 2 and 3 
clearly harden at higher luminosities (the brightest resolved \chan~source at the 
position of src. 3 is plotted, although the same is true for all 3 
resolved sources). This variety of behaviours is redolent of what we observe in 
lower-luminosity ULXs (\cf Kajava \& Poutanen 2009; Feng \& Kaaret 2009).

\section{Discussion}

Although X-ray data for this sample of ULXs is of limited
quality, a number of interesting insights can be drawn. 
The extreme sources in this sample tend to have spectra at the hard end of
the range observed in lower luminosity ULXs. This could be interpreted by
a direct comparison with Galactic black hole binaries (BHBs) as
being in a low/hard state, tending to support the IMBH hypothesis. However such an
interpretation of the X-ray spectral evidence may be na\"ive as it has been
shown that a sample of high quality ULX spectra reject such 
models (Gladstone 
\etal 2009). 
The success in fitting an absorbed power-law spectral model to the sample observations may 
be a reflection of the lack of data.

We find evidence of a high energy spectral break in 
src. 8. Such a spectral feature has been identified as the key signature 
of the proposed ultraluminous state in ULXs. Its presence in this highly luminous
source not only provides evidence against the low/hard state interpretation,
but further indicates that this extreme source may be drawn from the same population
as the less luminous sample of Gladstone \etal (2009). As the ultraluminous state 
likely occurs at much greater
Eddington ratios than are seen in the low/hard state, this negates the requirement for an 
IMBH, at least in this source.

The sample sources tend to be variable on short time-scales to a degree
greater than would be expected from linear scaling with flux from less
luminous ULXs. High levels of variability are seen in Galactic BHBs in
the low/hard state, 
however the notable increase in variability between `normal' ULXs and this 
sample may actually be attributable to the suppressed
variability identified in Heil \etal (2009) only being present in less luminous sources.

Similarly to lower luminosity ULXs, most of the sample sources reside in 
spiral galaxies. Src. 1 is observed to drop
in luminosity by a factor of $\sim 5$
from the hyperluminous regime to $\sim 1.9 \times 10^{40} 
{\rm erg s^{-1}}$. 
Such behaviour is not atypical of other well studied 
HLXs --  ESO 243-49 HLX 1 has been reported to 
vary in luminosity over $\sim$ days by a factor of $\sim 21$ (Godet \etal 2009), 
Wolter \etal (2006) report a factor of 2 dimming in the HLX candidate 
in the Cartwheel galaxy over 6 months and Ptak \& Griffiths (1999) report 
a factor of 4 reduction in the luminosity of M82 X-1. 

In summary there are indications of a number of similarities between the sample of highly
luminous ULXs presented here and less luminous sources - for example, 
there is a preference for spiral
galaxy hosts, and heterogenous spectral variability with luminosity is observed in both. 
Large changes in luminosity between observations are seen in the sample,
suggesting that the extreme brightnesses observed may be a transient phenomenon. 
Src. 3 in particular
decreases in luminosity such that it is indistinguishable from other less-luminous ULXs.
However, some differences are evident; unlike less luminous ULXs there is no evidence 
of supressed variability, or of a minority of spectrally soft sources (although 
with the latter point in particular this may be
due to the small sample size). Further, 
deeper observations of this interesting sub-class of ULXs are therefore required to improve our
understanding of these extraordinarily luminous objects.

\section{Acknowledgements}

We thank the anonymous referee for their extensive comments that have helped to much improve 
this paper. ADS and DJW acknowledge funding from the Science and Technology Facilities
Council in the form of a studentship. 
This work is based on observations obtained with \xmmn, an ESA science 
mission with instruments and contributions directly funded by ESA Member States and NASA.
It is also partially based on observations obtained with \chan, which is operated by NASA.

\end{document}